\documentclass[preprint,aps,prl,preprintnumbers,amsmath,amssymb,superscriptaddress,showpacs,floatfix]{revtex4}
\usepackage{epsfig}
\usepackage{dcolumn}
\usepackage{bm}
\usepackage{amssymb}
\begin{document}

\title{Thermal Width of the $\Upsilon$ at Large t' Hooft Coupling}
 
\preprint{BCCUNY-HEP/09-08}
\preprint{RBRC-798}

\author{Jorge Noronha}
\affiliation{Department of
Physics, Columbia University, 538 West 120$^{th}$ Street, New York,
NY 10027, USA}

\author{Adrian Dumitru}
\affiliation{Department of Natural Sciences, Baruch College, CUNY,
17 Lexington Avenue, New York, NY 10010, USA}
\affiliation{The Graduate School and University Center, City
  University of New York, 365 Fifth Avenue, New York, NY 10016, USA}
\affiliation{RIKEN BNL Research Center, Brookhaven National
  Laboratory, Upton, NY 11973, USA}

\begin{abstract}
We use the AdS/CFT correspondence to show that the heavy quark
(static) potential in a strongly-coupled plasma develops an imaginary
part at finite temperature. Thus, deeply bound heavy quarkonia states
acquire a small nonzero thermal width when the t'Hooft coupling
$\lambda=g^2 N_c \gg 1$ and the number of colors $N_c \to \infty$. In
the dual gravity description, this imaginary contribution comes from
thermal fluctuations around the bottom of the classical sagging string
in the bulk that connects the heavy quarks located at the boundary. We
predict a strong suppression of $\Upsilon$'s in heavy-ion collisions
and discuss how this may be used to estimate the initial temperature.
\end{abstract}

\maketitle

The conjectured equivalence of strongly-coupled 4-dimensional
$\mathcal{N}=4$ Supersymmetric Yang-Mills (SYM) to type IIB string
theory on AdS$_{5}\otimes S_{5}$ \cite{maldacena} has led to new
insight into the strong coupling dynamics of large $N_c$ gauge
theories at finite temperature. In fact, the Anti-de Sitter/Conformal
Field Theory (AdS/CFT) correspondence has been particularly useful to
compute real time correlators of gauge invariant quantities in
strongly-coupled plasmas such as 2-point functions of the
energy-momentum tensor at finite temperature \cite{miscAdS}. For
instance, it was shown that the shear viscosity to entropy density
ratio satisfies $4\pi\,\eta/s \geq 1$ in all strongly-coupled gauge
theories that possess a dual description in terms of supergravity
\cite{viscobound}. Here, we adopt strongly coupled $\mathcal{N}=4$ SYM
as a toy model for the deconfined, high temperature phase of QCD.

The dual description of the gauge theory at finite temperature
involves a near-extremal black brane in the bulk, which leads to a
5-dimensional metric (in real time) given by
\begin{equation}
\ ds^2 = -G_{00}(U)dt^2+G_{xx}(U)d\vec{x}^2+G_{UU}(U)dU^2~.
\label{metric}
\end{equation}
$G_{00}(U_h)=0$ defines the location $U_h$ of the black
brane horizon in the 5th coordinate, and the boundary is at
$U\to\infty$.

The potential between fundamental static sources separated by a
distance $L$ at large t' Hooft coupling $\lambda$ in $\mathcal{N}=4$
SYM was computed in \cite{Maldacena:1998im,Rey:1998ik} and shown to be
proportional to $1/L$ (due to conformal invariance of the theory) and
to $\sqrt{\lambda}$, which indicates that charges are partially
screened even in vacuum \cite{Maldacena:1998im,vacuumscreening}. In
general, thermal effects are expected to reduce the binding energies
of small states of very heavy quarks at high $T$. At strong coupling
thermal screening corrections appear at the same order in $\lambda$ as
the vacuum potential \cite{Brandhuber:1998bs}, as opposed to Debye
screening in weakly coupled quark-gluon plasmas
\cite{debyepqcd}. However, at distances $L < 1/T$ these corrections
are suppressed by a factor of $(LT)^4$, which originates from the
behavior of the dual bulk geometry near the black brane
horizon. Thermal effects also diminish as $\eta/s$
increases~\cite{Noronha:2009ia}.

In this Letter we show that at finite temperature the static potential
in a strongly-coupled plasma develops an imaginary part due to
fluctuations about the extremal configuration, which corresponds to a
string connecting the fundamental sources at the boundary of the
geometry. Such an imaginary part arises also in perturbative Quantum
Chromodynamics (pQCD) at order $\sim g^4$ due to Landau damping of the
static gluon exchanged by the heavy quark sources \cite{ImV_HTL}. At
large t' Hooft coupling, however, it appears already at the same order
as the vacuum potential, i.e., at ${\cal O}
(\sqrt{\lambda})$. Therefore, energy levels in this potential are not
sharp because they acquire a thermal width, $E=E_{\rm vac}+\Delta
E_T-i\Gamma$. The width $\Gamma$ is smaller than the vacuum energy
$E_{\rm vac}$ if $LT < 1$ (which is the relevant regime in the limit
of very heavy quarks, $m_Q\to\infty$) and of the same order in both
$\lambda$ and $LT$ as the shift $\Delta E_T$ of the real part of the
potential due to thermal screening effects. We propose that the
imaginary part of the potential mentioned above can be observed
experimentally via the suppression of $\Upsilon$ to dilepton decays
in heavy-ion versus p+p collisions at RHIC and LHC.

The relevant operator for our discussion is the path-ordered Wilson
loop defined as
\begin{equation}
\ W(C)=\frac{1}{N_c}{\rm Tr} \,P \,e^{i\int \hat{A}_{\mu}dx^{\mu}}
\label{wilsonloop}
\end{equation}
where $C$ denotes a closed loop in the boundary, $\hat{A}^{\mu}$ is
the non-Abelian gauge field, and the trace is over the fundamental
representation of $SU(N_c)$. We consider a rectangular loop with one
direction along the time coordinate $t$ and spatial extension $L$. In
the asymptotic limit $t\to \infty$, the vacuum expectation value of
the loop defines a static potential via $\langle W(C) \rangle \sim
e^{-i \,t \,\mathcal{V}_{Q\bar{Q}}(L)}$. This is what we call the ``heavy quark potential''.

The expectation value of $W(C)$ can be calculated at strong t'Hooft
coupling in non-Abelian plasmas at large $N_c$ that admit a weakly
coupled dual gravity description according to AdS/CFT
\cite{maldacena}. More specifically,
\begin{equation}
\ \langle W(C) \rangle_{CFT} = Z_{string}
\label{defineExpectWilson}
\end{equation}
where $Z_{string}$ is the full supersymmetric string generating
functional, which is defined in a 10 dimensional background spacetime
and includes a sum over all the string worldsheets whose boundary
coincide with $C$. In the supergravity approximation
$\lambda=g^2 N_c\,\gg 1$ and $N_c\to \infty$ and, in this case, an
infinitely massive excitation in the fundamental representation of
$SU(N_c)$ in the CFT is dual to a classical string in the bulk hanging down from a probe brane at infinity
\cite{Maldacena:1998im,Rey:1998ik}. Within this approximation
$Z_{string}\sim e^{i\,S_{NG}}$ and the dynamics of the string is given
by the classical Nambu-Goto (NG) action (we neglect the contribution
from other background fields such as the dilaton)
\begin{equation}
\ S_{NG}=-\frac{1}{2\pi \alpha'}\int d^2\sigma \sqrt{-{\rm det} \,h_{ab}}
\label{nambugotoaction}
\end{equation}
where $h_{ab}=G_{\mu\nu}\partial_{a}X^{\mu}\partial_{b}X^{\nu}$
($a,b=1,2$), $G_{\mu\nu}$ is the background bulk metric,
$\sigma^{a}=(\tau,\sigma)$ are the internal world sheet coordinates,
and $X^{\mu}=X^{\mu}(\tau,\sigma)$ is the embedding of the string in
the 10-dimensional spacetime. For $\mathcal{N}=4$ SYM, the configuration that minimizes the action is a
U-shaped curve that connects the string endpoints at the boundary and
has a minimum at some $U_*$ in AdS$_5$
\cite{Maldacena:1998im,Rey:1998ik}. 

The induced metric is
\begin{equation}
\ {\rm det}\,h_{ab}=X^{'\,2}\cdot \dot{X}^2-(\dot{X}\cdot X')^2
\label{detstring}
\end{equation}
where $X^{'\,\mu}(\tau,\sigma)=\partial_{\sigma}X^{\mu}(\tau,\sigma)$
and $\dot{X}^{\mu} (\tau,\sigma) = \partial_{\tau} X^{\mu}
(\tau,\sigma)$. We choose a gauge where the coordinates of the static
string are $X^{\mu}=(t,x,0,0,U(x))$, where $\tau=t$ and $\sigma =
x$. We neglect the string dynamics in the 5 dimensional compact space
and perform the calculation in real time at the boundary. In
fact, fixing the extremal configuration in such a way implies the
$t\to\infty$ limit; while a Wick rotation is of course still possible
(by switching to Euclidean metric, which would still give a complex
expectation value for the Wilson loop) one can no longer perform an
analytic continuation to imaginary time where the expectation value
should be real.

In this gauge,
\begin{equation}
\ S_{NG}=-\frac{\mathcal{T}}{2\pi \alpha'}\int_{-L/2}^{L/2} dx
\sqrt{U^{\prime\,2}+V(U)}~. 
\label{generalNG}
\end{equation}
where $\mathcal{T}\to\infty$ is the total time interval. Note that we
have assumed that $G_{00}G_{UU}=1$ (which is in general valid when $\lambda \to
\infty$) and defined $V(U)\equiv G_{00}G_{xx}$, which satisfies $V(U)\geq 0$ for $U\in [U_h,\infty)$. The equations of motion obtained
from Eq.\ (\ref{generalNG}) determine the classical string profile
$U_{c}(x)$ as discussed in detail in
refs.~\cite{Maldacena:1998im,Rey:1998ik,Brandhuber:1998bs,Noronha:2009ia}. The solution
$x=x(U_c)$ satisfies the following boundary condition
\begin{eqnarray}
 \frac{L}{2} &=& \int_{U_*}^{\infty} dU\,
 \left\{V(U)\left[\frac{V(U)}{V(U_*)}-1\right] \right\}^{-1/2}\;,
\label{profile1}
\end{eqnarray}
which is used to obtain $U_*=U_*(L,T)$. The expectation value of the
Wilson loop is obtained by substituting the classical string solution
$U_{c}(x)$ into the action in Eq.\ (\ref{generalNG}). In general, when
$LT \ll 1$ the dominant contribution to the potential comes from the
extremal worldsheet configuration described above and the potential is
computed as a series in $LT$. Other configurations are expected to
contribute significantly when $LT > 1$ \cite{Bak:2007fk}. However,
$L < 1/T$ is in fact the region of interest for bound states of very
heavy quarks which have small radii.

The heavy quark potential in the vacuum of $\mathcal{N}=4$ SYM has the
following simple analytical form (after subtracting the self-energy
contribution from the infinitely massive quarks)
\cite{Maldacena:1998im}
\begin{equation}
\ \mathcal{V}_{Q\bar{Q}}(L)= -\frac{4\pi^2}{\Gamma(1/4)^4}
\frac{\sqrt{\lambda}}{L}~,
\label{maldacenaprop}
\end{equation}
which may be compared to the standard SU($N_c$) Coulomb potential at
large $N_c$ corresponding to weak coupling:
\begin{equation}
\ \mathcal{V}_{\rm Coul}(L)= -\frac{1}{8\pi}
\frac{g^2 N_c}{L}~.
\label{Coulomb_largeN}
\end{equation}

We now take into account thermal fluctuations around the classical
solution $U_c(x)$. We shall show that the Wilson loop develops an
imaginary part due to fluctuations near the bottom of the classical
string configuration $U_*$. We consider long wavelength fluctuations of the string profile $U_c(x)
\to U_c(x) +\delta U(x)$ (with $\delta U'\to 0$), which give the
leading contribution to the string partition function in the
supergravity approximation as follows
\begin{equation}
\ Z_{string}\sim \int \mathcal{D}X^{\mu}\,e^{i\, S_{NG}(X)} \sim \int
\mathcal{D}\,\delta U(x)\, e^{i\, S_{NG}(U_c+\delta U)}\,.
\label{stringpartition1}
\end{equation}
When the bottom of the classical string is sufficiently close to the
horizon (though still above it), the worldsheet fluctuations, $\delta
U(x)$, near $x=0$ (where $U'_c=0$) can change the overall sign of the
argument of the NG square root and generate an imaginary contribution
to the action. In this case, both $U^{\prime\,2}_c$ and $V(U_c)$ are small and
the NG square root cannot be expanded in powers of $\delta U$.

The integral over $\delta U(x)$ is performed by dividing the $x$
interval in $2N$ parts such that $x_N=L/2$, $x_{-N}=-L/2$, $x_j=j\Delta
x$:
\begin{equation}
\ Z_{string}\sim  \int d(\delta U_{-N}) \cdots d(\delta U_N)\,
e^{-i\frac{\mathcal{T} \Delta x}{2\pi\alpha'} \sum_{j}
\sqrt{U_j^{\prime\,2}+V(U_j)}}~.\\
\label{stringpartition2}
\end{equation}
Near $x=0$ we expand $U_c(x_j) \simeq U_* + x_j^2 \,
U_c^{\prime\prime}(0)/2$ and $U_j^{\prime\,2}+V(U_j) \simeq
C_1x_j^2+C_2$, with
\begin{eqnarray}
C_1 &=& \frac{1}{2} U_c''(0) \left( 2U_c''(0)+V_*'+\delta U\, V_*'' +
\frac{1}{2} \delta U^2\, V_*'''\right) \nonumber\\
&\simeq& \frac{1}{2} U_c''(0) \left( 2U_c''(0)+V_*' \right)  \geq 0 \nonumber\\
C_2 &=& V_* + \delta U\, V_*' + \frac{1}{2} \delta U^2\, V_*'' ~.
\end{eqnarray}
Here, $V_*=V(U_*)$, $V^{\prime}_* = V'(U_*)$ and so on. The
imaginary part of the $Q\bar{Q}$ potential arises from the region of
$\delta U$ where $C_2<0$.

We isolate the contribution to the path integral from $x=x_j$,
\begin{equation}
\ I_j \equiv  \int_{\delta U_{j\,min}}^{\delta U_{j \, max}}
d(\delta U_j)\,
\exp\left\{-i\frac{\mathcal{T} \Delta x}{2\pi\alpha'} \left(x_j^2 C_1 +C_2
\right)^{1/2}\right\}\\
\label{nearx0contrib}
\end{equation}
where $\delta U_{j\, min,max}<0$ are defined as the zeros of the
argument of the square root. In this range of $\delta U_j$ the square
root in the exponent develops an imaginary part. (The complement of
the integration region in Eq.~(\ref{nearx0contrib}) provides a
correction to the real part of the potential due to fluctuations about
the extremal configuration that is not considered here.)  Note that in
this case $U_* + \delta U \sim U_h$, i.e., the bottom of the
fluctuating string touches the horizon. This may be viewed in the dual
gauge theory as a process analogous to Landau damping of the static
color fields which bind the quarks together, leading to the
formation of two unbound heavy quarks in the high-temperature plasma. 

On the other hand, at lower temperatures on the order of the QCD
crossover temperature and below, the dominant process should instead
correspond to the breakup $Q\bar{Q}\to (Q\bar{q})\, (\bar{Q}q)$ into
color-singlet heavy-light bound states; $q$ stands for a light
quark. Such tunnelling processes provide an exponentially small
thermal width of deeply bound states~\cite{KMcLS}. It should be clear
that the problem of quark tunneling cannot be solved rigorously since
it involves genuinely non-perturbative QCD dynamics. However, the
large mass of the heavy quark allows one to use the quasiclassical
approximation~\cite{KMcLS}. For a calculation of $Q\bar{Q}\to
(Q\bar{q})\, (\bar{Q}q)$ via the AdS/CFT correspondence see
Ref.~\cite{Cotrone:2005fr}. Here, too, the temperature must be
sufficiently low to allow for the formation of two new heavy-light
quark bound states. In the gravity description, the final state
corresponds to two strings connecting the Q-brane with the q-brane
while in our approach they connect to the black-hole horizon.

The factor $1/\alpha'$ in the exponent implies that the leading order
contribution is of order $\mathcal{\sqrt{\lambda}}$. In the
supergravity approximation $\lambda \gg 1$ and, thus, $I_j$ can be
computed in the saddle-point approximation. This gives $\delta U = -
V^{\prime}_* / V^{\prime\prime}_*$ and so
\begin{eqnarray}
\ \exp\left\{ -i \,\mathcal{T} \,\mathcal{V}_{Q\bar{Q}} \right\} 
&=& \prod\limits_j I_j \nonumber\\
& & \hspace*{-3cm}
\sim  \exp\left\{-\,\frac{\mathcal{T}}{2\pi\alpha'}
\left[\int_{|x|<x_c}dx \,
  \sqrt{-x^2 C_1 - V_*+V^{\prime\,2}_*/2V^{\prime\prime}_*}
  \right.\right. \nonumber\\
& +&\left.\left. i \int_{|x|>x_c}dx \, \sqrt{U_c^{\prime\,2}+V(U_c)}\right]
\right\}
\label{nearx0contrib2}
\end{eqnarray}
where $x_c = \sqrt{(- V_* + V^{\prime\,2}_* / 2V^{\prime\prime}_*) /
  C_1}$ if this root is real, and $x_c=0$ otherwise. The second
contribution in the exponent gives the real part of the $Q\bar{Q}$
potential which we drop from now on (see
refs.~\cite{Maldacena:1998im,Rey:1998ik,Brandhuber:1998bs,Noronha:2009ia}). 
Performing the integral over $|x|<x_c$ we find
\begin{equation}
 \,\mathrm{Im}~\mathcal{V}_{Q\bar{Q}} =
 -\frac{1}{2\sqrt{2} \alpha'}
\left[\frac{V^{\prime}_*}{2V^{\prime\prime}_*}-\frac{V_*}{V^{\prime}_*}
\right]\,,
\end{equation}
where we used that $U_c^{''}(0)=V^{\prime}_*/2$. For $\mathcal{N}=4$
SYM at large $\lambda$ the potential entering the NG action is given by
$V(U)=\left(U^4-U_h^4\right)/R^4$, where $R$ is the radius
of AdS$_5$ and $U_h=\pi R^2 T$. This gives
\begin{equation}
\ \, \mathrm{Im}~\mathcal{V}_{Q\bar{Q}} = -\frac{\pi}{24\sqrt{2}}\,
\sqrt{\lambda}\,T \, \frac{3\zeta^4-1}{\zeta} \,,
\label{nearx0contrib3}
\end{equation}
where $\sqrt{\lambda}=R^2/\alpha'$ and $\zeta\equiv U_h/U_*<1$. Note
that the equation above applies only when $\zeta >3^{-1/4}\approx
0.76$; otherwise $\mathrm{Im}~\mathcal{V}_{Q\bar{Q}}=0$ because the
solution for $x_c$ from above ceases to exist. Thus, in the vicinity
of this point the width generated by the imaginary part of the
potential (see below) is small compared to the binding energy.

The dependence of $\zeta$ on $L$ and $T$ can be found from
Eq.\ (\ref{profile1}). At small $LT$ we have $LT=b\,\zeta$ with
$b=2\Gamma(3/4)/\sqrt{\pi}\,\Gamma(1/4)\approx 0.38$. On the other hand,
when the bottom of the classical string comes too close to the
horizon, $\zeta\sim 0.85$, the U-shaped configuration
used here receives higher-order corrections \cite{Bak:2007fk}
and cannot be used anymore. With $\zeta\sim LT$,
the imaginary part of the potential~(\ref{nearx0contrib3})
is smaller than the dominant contribution to the real part,
eq.~(\ref{maldacenaprop}), by a factor $\sim (LT)^4$. Here, we
consider only temperatures such that thermal screening
corrections to the {\em real part} of the potential are small;
the bound state then probes the potential only in the region $LT<1$. 

The imaginary part in Eq.\ (\ref{nearx0contrib3}) shifts the Bohr
energy level obtained with the Coulomb-like vacuum potential~(\ref{maldacenaprop}),
$E_0\to E_0 -i\Gamma$; to first order,
\begin{eqnarray}
\Gamma_{Q\bar{Q}} &\equiv& -\langle \psi|
\mathrm{Im}~\mathcal{V}_{Q\bar{Q}} |\psi\rangle \nonumber\\
&=&
\frac{\pi\sqrt{\lambda}}{48\sqrt{2}}\frac{b}{a_0}
 \left[45\left(\frac{a_0T}{b}\right)^4 -2\right]~,
\label{width}
\end{eqnarray} 
where $|\psi\rangle$ denotes the unperturbed Coulomb ground state wave
function and $a_0=\Gamma(1/4)^4/2\pi^2\sqrt{\lambda}\,m_Q$ is the Bohr
radius. The width decreases with the quark mass and with the t' Hooft
coupling, approximately as $\Gamma_{Q\bar{Q}}\sim 1/\lambda\, m_Q^3$;
it increases rapidly with the temperature, $\sim T^4$. For
$m_Q=4.7$~GeV, $T=0.3$~GeV, $\sqrt\lambda=3$~\cite{Noronha:2009vz} we
obtain $\Gamma_\Upsilon \simeq 48$~MeV.

It is interesting to compare this result to the imaginary part of the
heavy quark potential computed in pQCD:
$\mathrm{Im}~\mathcal{V}_{Q\bar{Q}} \sim -
\alpha_s^2 C_F N_c T (LT)^2\log\, (LT)^{-1}$ \cite{ImV_HTL}. This is
smaller than $\mathrm{Re}~\mathcal{V}_{Q\bar{Q}} \sim \alpha_s C_F/L$ by one
power of the coupling and three powers of $LT$. With $L\sim (\alpha_s
C_F m_Q)^{-1}$, the width decreases less rapidly with the quark mass
than for $\mathcal{N}=4$ SYM at strong coupling. However, the
numerical value of $\Gamma_\Upsilon$ is on the order of tens of
MeV, similar to what we obtain here.

We suggest that the width computed above is accessible experimentally
through the suppression of $\Upsilon\to \ell^+ \ell^-$ processes in
heavy-ion collisions at RHIC or LHC. Neglecting ``regeneration'' of
bound states from $b$ and $\bar{b}$ quarks in the medium we can
estimate the number of $\Upsilon$ mesons in the plasma at mid-rapidity,
which have not decayed into unbound $b$ and $\bar{b}$ quarks up to
time $t$ after the collision, from
\begin{eqnarray}
\frac{dN}{dt} &=& -\Gamma_\Upsilon(T(t))\,N(t) \nonumber \\ \rightarrow 
N(t) &\simeq& N_0 \exp\left(-\int dt\, \Gamma_\Upsilon(t)\right)~.
\end{eqnarray}
This solution assumes that $\Gamma_\Upsilon(T(t))$ is a slowly varying
function of time. The initial number of $\Upsilon$ states may be
estimated from the multiplicity in p+p collisions times the number of
binary collisions at a given impact parameter: $N_0\simeq N_{\rm coll}
N_{pp}^\Upsilon$. Thus, the integrated ``nuclear modification factor''
$R_{AA}$ for the process $\Upsilon\to \ell^+ \ell^-$ is approximately
given by $R_{AA}(\Upsilon\to \ell^+ \ell^-) \simeq
\exp(-\bar\Gamma_\Upsilon \, t)$, where $\bar\Gamma$ denotes a
suitable average of $\Gamma(T)$ over the lifetime of the quark-gluon
plasma. Due to the strong temperature dependence of the width, this
average is dominated by the early stage and thus we expect that
$\bar\Gamma$ provides an estimate of the initial temperature in
heavy-ion collisions via Eq.\ (\ref{width}). For $t=5$~fm/c and
$\bar\Gamma_\Upsilon = 48$~MeV we obtain $R_{AA}\simeq 0.3$. An
experimental estimate for the thermal quarkonium decay rate
$\bar\Gamma$ could be obtained once a statistically significant
detection of the $\Upsilon\to \ell^+\ell^-$ process has been
achieved~\cite{data}.

We thank W.~Zajc for useful comments on the preprint of this manuscript.
J.N.\ acknowledges support from US-DOE Nuclear Science Grant No.\
DE-FG02-93ER40764. A.D.\ gratefully acknowledges support from The City
University of New York through a PSC-CUNY Research Award Program
grant and by the Office of Nuclear Physics, US-DOE Grant No.\
DE-FG02-09ER41620. J.N.\ and A.D.\ also thank Goethe University for
their hospitality and the Helmholtz International Center for FAIR
for support within the LOEWE program.

\end{document}